# INFLUENCE OF THE SUBSTRATE AND PRECURSOR ON THE MAGNETIC AND MAGNETO-TRANSPORT PROPERTIES IN MAGNETITE FILMS

Enio Lima Jr.[a,1], Giancarlo E. S. Brito[b], Christian Cavelius[c], Vladimir Sivakov[c], Hao Shen[c], Sanjay Mathur[c], Gerardo F. Goya[d]

[a] CONICET, Centro Atómico Bariloche, 8400 S. C. de Bariloche, RN, Argentina

[b] Universidade de São Paulo, Instituto de Física, CP 66318, São Paulo, BR-05315970 Brazil

[c] Leibniz Inst. New Mat., CVD Division, Saarbrucken, D-66123 Germany

[d] Instituto de Nanociencia de Aragon (INA), University of Zaragoza, Spain.

[1] Corresponding author: Phone: (54) 294 4445158 , fax: (54) 294 4445299, email: lima@cab.cnea.gov.ar

*ABSTRACT*

We have investigated the magnetic and transport properties of nanoscaled $Fe_3O_4$ films obtained from Chemical Vapor Deposition (CVD) technique using $[Fe^{II}Fe_2^{III}(OBu^t)_8]$ and $[Fe_2^{III}(OBu^t)_6]$ precursors. Samples were deposited on different substrates (*i.e.*, MgO (001), $MgAl_2O_4$ (001) and $Al_2O_3$ (0001)) with thicknesses varying from 50 to 350 nm. Atomic Force Microscopy analysis indicated a granular nature of the samples, irrespective of the synthesis conditions (precursor and deposition temperature, $T_{pre}$) and substrate. Despite the similar morphology of the films, magnetic and transport properties were found to depend on the precursor used for deposition. Using $[Fe^{II}Fe_2^{III}(OBu^t)_8]$ as precursor resulted in lower resistivity, higher $M_S$ and a sharper magnetization decrease at the Verwey transition ($T_V$). The temperature dependence of resistivity was found to depend on the precursor and $T_{pre}$. We found that the transport is dominated by the density of antiferromagnetic antiphase boundaries (AF-APB's) when $[Fe^{II}Fe_2^{III}(OBu^t)_8]$ precursor and $T_{pre}$ = 363 K are used. On the other hand, grain boundary-scattering seems to be the main mechanism when $[Fe_2^{III}(OBu^t)_6]$ is used. The Magnetoresistance (MR(H)) displayed an approximate linear behavior in the high field regime ( H > 796 kA/m), with a maximum value at room-temperature of ~ 2-3 % for H = 1592 kA/m, irrespective from the transport mechanism.

## 1. INTRODUCTION

Magnetite ($Fe_3O_4$) is a ferrimagnet below its Curie temperature $T_C = 850$ K, having an inverse spinel cubic structure with $O_{h7}$ (Fd3m) space group [1]. The Fe ions are distributed among two different crystallographic sites: the octahedral B site occupied by both $Fe^{2+}$ and $Fe^{3+}$ ions, and the tetrahedral A site, where only $Fe^{3+}$ ions are present. Previous works on band calculations in bulk magnetite have shown that only one of the spins channels has a gap at the Fermi level, suggesting that conduction electrons are fully spin-polarized in $Fe_3O_4$ [2]. In spite of the fact that this theoretical polarized band structure should display a large magnetoresistance (MR) effect, this consequence has not yet been observed experimentally. Indeed, only a modest MR effect has been so far observed at room temperature (up to 3-4 % in fields of 1592 kA/m) for thin films, whereas for high-quality $Fe_3O_4$ single crystals the MR effect is absent [3-6].

$Fe_3O_4$ films show divergences regarding the magnetic behavior as compared with bulk material. For example, Arora *et al* [7] have observed that saturation magnetization of epitaxial films is strongly related to the thickness of the film, obtaining an $M_S$ value of 1000 kA/m for 5 nm in comparison to the 512 kA/m of the bulk material [8]. Saturation field of epitaxial or polycrystalline films (up to 1194 kA/m) exceeds by far the corresponding value for single crystals (~ 24 kA/m) [9]; however, the experimental values on MgO obtained for the effective magnetic anisotropy on these films have explained assuming only magnetocrystalline and shape anisotropies, without any additional mechanism [10].

There are at present no general consensus regarding the detailed magnetic structure of $Fe_3O_4$ epitaxial films obtained from different synthesis routes such as molecular beam epitaxy (MBE), pulsed laser deposition (PLD) *etc.*, and also related to different substrates such as MgO, $Al_2O_3$, $MgAl_2O_4$, $BaTiO_3$ and $SrTiO_3$ [7,9-16], despite some results indicates that the magnetism and magneto-transport phenomena in this kind of film are controlled by the domain boundaries and the anti-ferromagnetic couple strength in the boundary [9]. For example, the close structural match between MgO and $Fe_3O_4$ cell parameters (about 0.3 %) usually yields to epitaxial growth on this substrate. However, it has been demonstrated that the lower symmetry and larger unit cell of the magnetite crystal structure can result in 'broken' cation sublattices at the $Fe_3O_4$ layers on MgO, as well as different directions on nucleation of $Fe_3O_4$ islands on the first stages of the growth process [10]. These considerations have been the basis of one of the key concepts to understand the magnetotransport properties of $Fe_3O_4$ epitaxial films through the antiphase boundary (APB). The APB is a natural growth defect in epitaxial films, whose origin is associated to discontinuities in the cation B sublattice, along definite directions on the angle and distances governing the A-A, B-B, and A-B exchange interactions. The magnetic coupling over a large fraction of these APB´s is antiferromagnetic (AF), yielding a local magnetic structure different than the bulk material [17]. It has been proposed that these AF-APBs are efficient scattering centers for the fully spin-polarized electrons that results in the increased resistance observed in thin films. The domain size or the APB density seems to be dependent of the thickness [17] and of the misfit with the structure of the substrate [15].

Polycrystalline $Fe_3O_4$ films were prepared by sputtering, chemical vapor deposition (CVD), PLD, electrodeposition, etc, on different substrates ($MgAl_2O_3$, $Al_2O_3$, quartz, glass, SiO, etc.) [9,18-20]. The structure and the granulometry for these films depends strongly from the synthesis method and growth conditions, as well as from the substrate. In this way, the magnetic and transport properties are strongly influenced by the synthesis method and growth conditions. In general, polycrystalline films present high saturation field as consequence of the grain boundary, and its value depend on the morphology, stoichiometry and structure of the film [9,18]. Resistivity is higher than the bulk one, as consequence of the scattering in the grain boundaries [9,18]. Magneto-resistive behavior are also observed for polycrystalline films and it is determined by the spin-polarized electron tunneling trough the grain boundary [9,18]. Magnetic and transport properties of $Fe_3O_4$ films are also dependent of crystallinity and stoichiometric characteristics, which are associated to the growth conditions too. For example, Bohra *et al* [19] have observed significant improves in the crystallinity and in the magnetic properties of annealed polycrystalline films, while Mantovan *et al* [18] correlate the number of vacancies and the stoichiometric characteristics with the magnetic and transport properties of the system.

From the above discussion it is clear that precise control over the morphology and phase purity is necessary to disentangle the underlying transport mechanisms, inasmuch as they depend strongly dependent on their chemical composition and microstructure. In this context, CVD processes using single molecular precursors offers a convenient method for the size- and morphology-controlled deposition of metal oxide film [21,22]. Molecular chemical compounds that mimic the atomic composition or bonding features of solid phases are attractive precursors because they allow a control over the evolution of materials from discrete single- or poly-atomic units (molecule) to the extended frameworks (bulk) [23-25]. From this, we can obtain a molecular design for conservation of valence and stoichiometry in CVD deposited Magnetite films. The mixed-valent iron alkoxide $[Fe^{II}Fe_2^{III}(OBu^t)_8]$ is a precursor that contains both Fe(II) and Fe(III) centers in a single framework. Thus, it is expected that it governs the formation of $Fe_3O_4$ by imposing a positional control on phase-building ions.

In this work, we have investigate the magnetic and magneto-transport properties of $Fe_3O_4$ films deposited by CVD with using two distinct precursors: $[Fe^{II}Fe_2^{III}(OBu^t)_8]$ and $[Fe_2^{III}(OBu^t)_6]$. The films were deposited in three different substrates ($MgO(001)$, $MgAl_2O_4$ (001) and $Al_2O_3$ (0001)). We have observed a strong and straight influence of the precursor on the magnetic and transport properties of $Fe_3O_4$ films, overlapping the influence of the mismatch between the structure of films and the substrates. These results are important for future application of precursors with molecular design for conservation of valence and stoichiometry for CVD-deposited Magnetite films.

## 2. MATERIALS AND METHODS

The molecular framework of the $[Fe^{II}Fe_2^{III}(OBu^t)_8]$ is formally constituted by a $Fe^{II}$ cation coordinated by two bidentate $\{Fe^{III}(OBu^t)_4\}^{-1}$ anions. All the iron atoms are present in a distorted tetrahedral environment of oxygen. This compound represents mixed-valent iron alkoxide mimicking the features of a mixed-valent condensed phase. $[Fe^{II}Fe_2^{III}(OBu^t)_8]$ is volatile and can be transferred intact in the vapor phase at 263 K/$10^{-3}$ Torr. The films properties of the films deposited from this precursor are compared with those deposited from the precursor $[Fe^{III}_2(OBu^t)_6]$, that contains only $Fe^{III}$ ions.

A total of nine $Fe_3O_4$ films were synthesized in a cold-wall Chemical Vapor Deposition (CVD) reactor using the iron alkoxide precursors as a single-source for $Fe^{II}$ and $Fe^{III}$ ions. The series was composed of three sets of three crystalline films deposited on (001) oriented MgO (mismatch ~0.3 %) and $MgAl_2O_4$ (mismatch ~3.9 %), as well as on (0001) oriented $Al_2O_3$ (mismatch ~8%). Each of these three sets were obtained for different synthesis conditions (systematically changing the temperature of the precursor $T_{pre}$ = 354 - 363 K, and deposition time), and using $[Fe^{III}_2Fe^{II}(OBu^t)_8]$ and $[Fe^{III}_2(OBu^t)_6]$ as precursors.

Table I summarizes the information about synthesis conditions for each sample. The temperature of the substrate ($T_{sub}$) was maintained the same (723 K) for all samples.

To further characterize the samples, we performed Rutherford Backscattering Spectroscopy (RBS) measurements in order to extract the thickness (d) and the composition of the films. The obtained film thicknesses, d, ranged from 50 to 350 nm whereas the relative amounts of Fe and O fitted from the profiles are close to the fully stoichiometric magnetite $Fe_3O_4$. Information about the thickness and composition obtained from RBS analysis of each sample is given in Table I.

TABLE 1. Synthesis conditions and the results from RBS and XRD analyzes for each sample. $T_{sub}$ = temperature of the substrate; $T_{pre}$ = temperature of the precursor; DPT = Deposition Time; d = thickness of the film obtained by RBS. Fe% and O% are the atomic mass percentage as obtained from RBS profiles.

| Sample | Substrate | Precursor | $T_{sub}$ (K) | $T_{pre}$ (K) | Deposition time | RBS d (nm) | Fe% | O% |
|---|---|---|---|---|---|---|---|---|
| S1 | $MgAl_2O_4$ (100) | $Fe_3(OBu^t)_8$ | 723 | 363±2 | 15 min. | 214 | 44 | 56 |
| S2 | MgO (100) | | | | | 200 | 38 | 62 |
| S3 | $Al_2O_3$ (0001) | | | | | 208 | 41 | 59 |
| S4 | $MgAl_2O_4$ (100) | $Fe_2(OBu^t)_6$ | 723 | 354±1 | 60 min. | 114 | 41 | 59 |
| S5 | MgO (100) | | | | | 71 | 44 | 56 |
| S6 | $Al_2O_3$ (0001) | | | | | 50 | 44 | 56 |
| S7 | $MgAl_2O_4$ (100) | $Fe_3(OBu^t)_8$ | 723 | 357±2 | 15 min. | 305 | 39 | 61 |
| S8 | MgO (100) | | | | | 250 | 38 | 52 |
| S9 | $Al_2O_3$ (0001) | | | | | 360 | 42 | 58 |

X-ray diffraction patterns were collected with θ-2θ geometry and using Ni-filtered Cu-$K_α$ radiation (λ=0.15418 nm). The resistivity measurements as a function of temperature (ρ(T)) were made using a DC four-probe method. For all contacts we obtained linear IxV curves at room temperature applying DC voltages (V) and measuring the current (I) in a four-point geometry. Magnetization curves were made in a commercial SQUID magnetometer as function of temperature (M(T)), in zero-field-cooling (ZFC) and field-cooling (FC) modes, and applied field (M(H)) up to 5570 kA/m, with applied field parallel (in-plane) and perpendicular (out-of-plane) to the film plane. Magneto-resistance curves (MR(H)) at room temperature were collected up to 1592 kA/m using a four-probe geometry, with applied field in- and out-plane. The magneto-resistance MR(H) was calculated using the relationship:

$$MR(H) = (R(H) - R(0))/R(0), \qquad \text{eq. (1)}$$

where R(0) is the resistance at zero applied magnetic field. Magnetic (MFM) and Atomic force microscopy (AFM) images were made in Nanoscope III A – Digital Instruments, operating in tapping mode, and phase contrast for MFM images.

# 3. RESULTS AND DISCUSSION

Table I shows a difference in the deposition rate of the films depending of the precursor and $T_{pre}$, as calculated by the thickness from RBS analysis. The highest deposition rate were obtained for the set formed by samples S7, S8 and S9 (deposited from precursor $[Fe^{II}Fe^{III}{}_2(OBu^t)_8]$ at $T_{pre}=$ 357 K), being around 16.7 – 24.0 nm/min. Close deposition rates were observed for samples S1, S2 and S3 (13.3 – 14.3 nm/min.), also deposited from precursor $[Fe^{II}Fe^{III}{}_2(OBu^t)_8]$, but at $T_{pre}=$ 363 K. Finally, the set formed by samples S4, S5, and S6, which were deposited from precursor $[Fe^{III}{}_2(OBu^t)_6]$ at ($T_{pre}$ = 354 K), presents the lowest deposition rates (0.8 – 1.9 nm/min.). Therefore, the deposition kinetic drastically changes with the precursor used in the process, while the effects of the temperature of the precursor $T_{pre}$ are significantly smaller. Table I also gives the composition values obtained from RBS, being between 38 % and 44 % *at.* Fe for all samples. The nominal value for the bulk magnetite is ~ 42.8 % *at.* Fe.

Crystallographic analysis (XRD) of sample S7 (Figure 1-a), deposited from precursor $[Fe^{II}Fe^{III}{}_2(OBu^t)_8]$ at $T_{pre}=$ 357 K, shows the peak characteristic of (004) plane of magnetite together with the intense peak corresponding to the (001) direction of the $MgAl_2O_4$ precursor. We also observe the peak relative to the direction (311) with very low intensity in comparison to the (004) one (about 50 times smaller). The main contribution to the XRD profile of this sample can be fitted with three pseudo-voigt curves: peak (004) of magnetite and the $K\alpha_1$ and $K\alpha_2$ contributions of the peak (001) from the substrate. Full width at half maximum (FWHM) for the $Fe_3O_4$ (004) peak is 0.8º against a FWHM= 0.1º for the $MgAl_2O_4$ (004) peak. For comparison, epitaxial films of $Fe_3O_4$ with roughness about 0.28 nm deposited by PLD on $MgAl_2O_4$ show FWHM ~ 0.2º - 0.3º for the (004) peak [26,27] while the FWMH= 1º was observed for polycrystalline film growth by CVD on $MgAl_2O_4$ [18]. Similar characteristics were exhibited by the the XRD profiles of samples S8 (precursor $[Fe^{II}Fe^{III}{}_2(OBu^t)_8]$, $T_{pre}=$ 357 K) and S5 (precursor $[Fe^{III}{}_2(OBu^t)_6]$, $T_{pre}=$ 354 K) deposited on MgO (001). Figure1-b presents the XRD profile of sample S9, deposited on $Al_2O_3$ (0001), with the precursor $[Fe^{II}Fe^{III}{}_2(OBu^t)_8]$ and $T_{pre}=$ 357 K, where the diffraction lines (311), (222), (004), (422) and (511) of magnetite are observed together with the diffraction line (0001) of $Al_2O_3$, although the magnetite peaks present distinct intensity relation when compared with that one expected for bulk material. According to this XRD analysis, the films deposited from the precursor $[Fe^{II}Fe^{III}{}_2(OBu^t)_8]$ are very crystalline, with the films on MgO and $MgAl_2O_4$ being strongly oriented in the (004) direction, while the films on $Al_2O_3$ present several crystalline orientations.

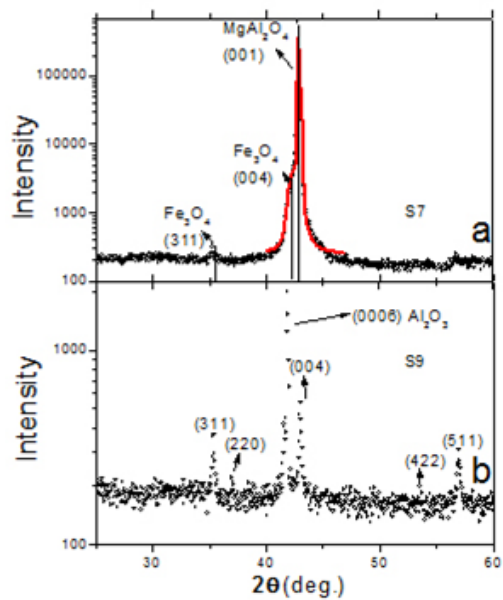

Figure 1. X-ray diffraction (XRD) patterns of films (a) S7 and (b) S9 grown on $MgAl_2O_4$(001) $Al_2O_3$, respectively. Solid line is the fitting with 3 pseudo-Voigt curves corresponding to the peaks of $Fe_3O_4$ (004) direction together with $K_{\alpha1}$ and $K_{\alpha2}$ peaks of the (001) direction of $MgAl_2O_4$.

Figure 2-a-c present the AFM images of samples deposited from both precursors on MgO (S2, S5 and S8), evidencing the granular nature of these films, with rms = 8.2, 7.9 and 4.1 nm, respectively. AFM images of samples S7 and S9, deposited from precursor [$Fe^{II}Fe^{III}_2(OBu^t)_8$] on $MgAl_2O_4$ and $Al_2O_3$, respectively, are presented in figure 2-d and -e, also showing a granular nature and rms = 8.0 and 9.5 nm, respectively.

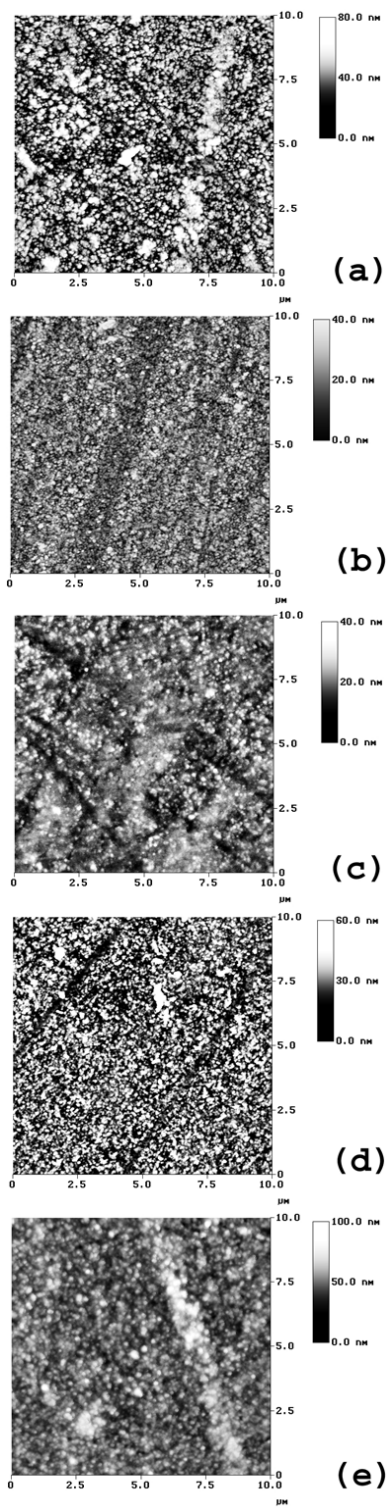

Figure 2. AFM images of samples (a) S2, (b) S5 and (c) S8 grown on MgO (001) from both precursors, and Figure 2-d and -e present the AFM images of samples S7 and S9, which were grown on MgAl$_2$O$_4$ and Al$_2$O$_3$, respectively, from precursor [Fe$^{II}$Fe$^{III}$$_2$(OBu$^t$)$_8$].

Correlating the AFM and XRD results, all films prepared from precursor [Fe$^{II}$Fe$^{III}$$_2$(OBu$^t$)$_8$] present a granular nature and high crystallinity, with the films growth on MgO and MgAl$_2$O$_4$ presenting a preferential growth on the (004) direction of the magnetite structure, while the film grown on Al$_2$O$_3$ does not present a preferential growth direction. These results are probably related to the structural mismatch between the Fe$_3$O$_4$ films and the different substrates: for the mismatch with MgO (0.3 %) or MgAl$_2$O$_4$ (3.9 %) is smaller than Al$_2$O$_3$ (~ 8.9 %). At the same time, AFM and XRD results indicates that the granular nature of the films and the crystallinity present no dependence with the precursor or the T$_{pre}$ used in the deposition procedure.

*3.1 - Magnetization*

Figure 3 presents the M(T) curves (H= 798 A/m) measured at ZFC and FC modes for all samples. The Verwey Transition (temperature of charge ordering - T$_V$) is clearly evidenced in all curves as a sharp drop in the magnetization for both curves. T$_V$ was assumed as the maximum value of the ZFC derivative curve and varies between 110-118 K. The values of T$_V$ and the width of the transition in the temperature axis (δT$_V$) of each sample are given in table II. For comparison, we also present in figure 3 the M(T) curves of a commercial magnetite monocrystal, which shows a markedly (δT$_V$= 8 K) Verwey transition at T$_V$ = 109 K. This value of T$_V$ is lower than the expected for the magnetite (122 – 125 K), indicating a variation in the stoichiometry of the monocrystal with respect to magnetite [2]. T$_V$ values obtained for the films are between the monocrystal and the bulk ones, at the same time, our values of δT$_V$ are slightly higher than that one of the monocrystal. These are evidences of the stoichiometric and structural quality of our films, since T$_V$ is strongly affected by these factors. In figure 4, we present the T$_V$ as function of the thickness, showing clearly three groups of samples: (S1, S2, S3), (S4, S5, S6) and (S7, S8, S9). These groups are reflected in the plot of δT$_V$ *vs.* thickness (see inset of figure 4) too, except for sample S9, closer to samples S1, S2, and S3. Therefore, there is a straight connection between the synthesis conditions (precursor and T$_{pre}$) and T$_V$ and δT$_V$. The group (S1, S2, S3) (deposited from [Fe$^{II}$Fe$^{III}$$_2$(OBu$^t$)$_8$] with T$_{pre}$ = 363 K) presents the smaller values of δT$_V$. Group (S7, S8, S9) have intermediate values of δT$_V$ (same precursor with T$_{pre}$= 357 K), while the group (S1, S2, S3) present the highest ones (precursor [Fe$^{III}$$_2$(OBu$^t$)$_6$] and T$_{pre}$= 354 K).

Figure 5-a shows the M(H) curves for in-plane direction of all samples measured at room temperature and the magnetization values are given in kA/m by using the volume of the film in each sample. The diamagnetic component of the corresponding substrate was subtracted from the M(H) curves. It is not observed significant differences in the coercive field H$_C$ = 14 - 15 kA/m of the samples. However, it is clear a difference in the saturation magnetization (M$_S$). As observed in table II, samples prepared from precursor [Fe$^{II}$Fe$^{III}$$_2$(OBu$^t$)$_8$] (samples S1-S3 and S7-S9) have M$_S$ values between 518 and 558 kA/m, while samples from precursor [Fe$^{III}$$_2$(OBu$^t$)$_6$] (samples S4-S6) present no saturation up to 1592 kA/m, with extrapolated values given M$_S$ = 450 – 467 kA/m. The absence of saturation for the fields used in our magnetization measurements is expected for polycrystalline films as consequence of the grain boundaries [28]. In-plane and out-of-plane M(H) curves of sample S5 and S8 are compared in figure 5-b, showing that the in-plane in the easy direction for both (demagnetization factor was taken into account in the out-of-plane curves). This result is expected because of the shape anisotropy, with in-plane and out-of-plane curves converging one to another at H = 398 kA/m, which indicates the intensity of the shape anisotropy.

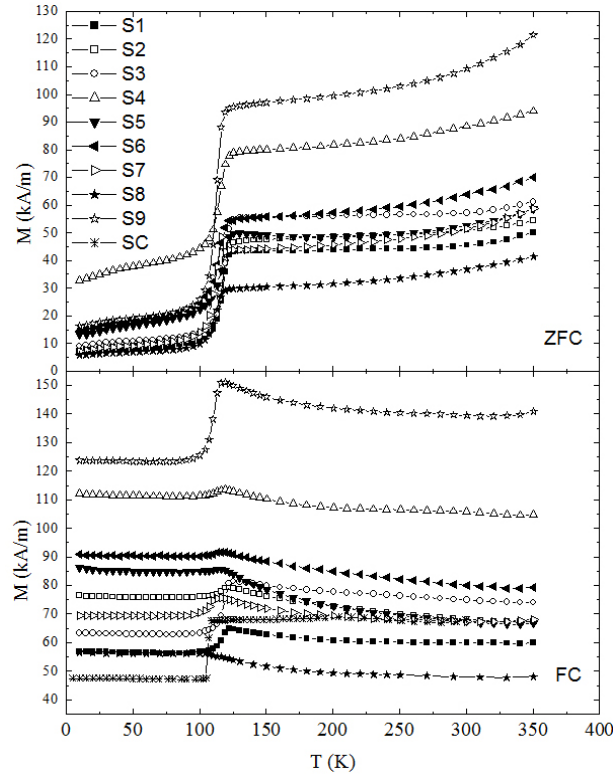

Figure 3. M(T) curves (H = 796 kA/m) measured at ZFC and FC modes for all samples. For comparison, we also present in figure 3 the M(T) curves of a commercial magnetite monocrystal.

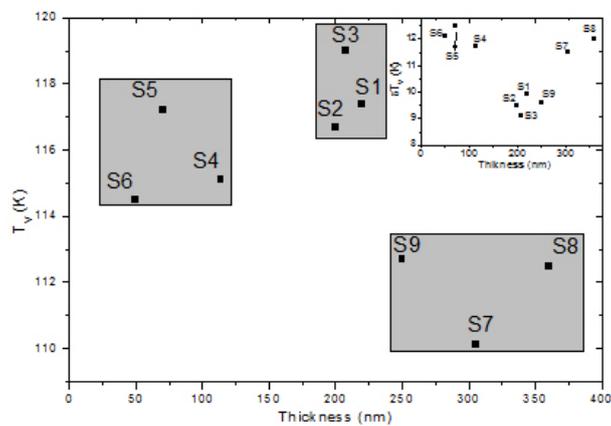

Figure 4. Temperature of the Verwey Transition ($T_V$) as function of the film thickness, showing clearly three groups of samples: (S1, S2, S3), (S4, S5, S6) and (S7, S8, S9). Inset: plot of $\delta T_V$ vs. thickness of all samples.

In spite of the same granular nature of all samples in the present series, some differences in their magnetic properties were observed in their values of $M_S$, saturation field, $T_V$ and $\delta T_V$. Regarding $M_S$ values, the highest values were observed for those samples prepared from [$Fe^{II}Fe^{III}_2(OBu^t)_8$] precursor. Concurrently, those samples were found to saturate at lower applied fields. We have previously mentioned that a direct relation between the precursor and $T_{pre}$ with $T_V$ and $\delta T_V$ was observed. We propose that this dependence is associated to the deposition kinetics (thickness – see fig. 4): group (S1, S2, S3) (deposited from [Fe$^{II}$Fe$^{III}_2$(OBu$^t$)$_8$] with T$_{pre}$ = 363 K) presents the smaller values of δT$_V$; group (S7, S8, S9) have intermediate values of δT$_V$ (same precursor with T$_{pre}$= 357 K); and group (S1, S2, S3) (precursor [Fe$^{III}_2$(OBu$^t$)$_6$] and T$_{pre}$= 354 K) present the highest values of T$_{pre}$ and δT$_V$. These results indicate that the precursor [Fe$^{II}$Fe$^{III}_2$(OBu$^t$)$_8$] produce samples with high magnetic quality in comparison to the precursor [Fe$^{III}_2$(OBu$^t$)$_6$], probably as consequence of the higher local crystallographic order in the samples associated to the first one, which is reflected in the values of T$_{pre}$ and δT$_V$. The differences in the magnetic properties between samples prepared from the two precursors is also probably related to the differences in the deposition kinetic, as evaluated from RBS analysis (see Table I).

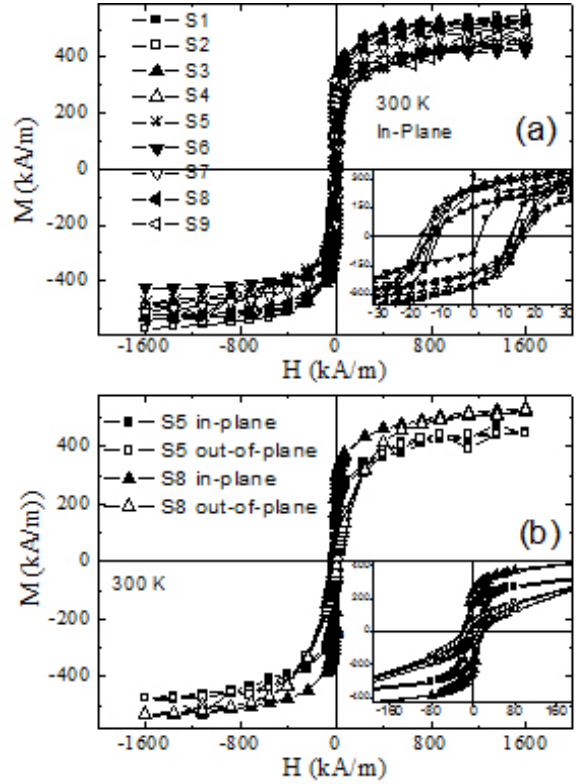

Figure 5. Figure 5-a shows the M(H) curves for in-plane direction of all samples measured at room temperature and the magnetization values are given in kA/m by using the volume of the film in each sample. Figure 5-b displays the in-plane and out-of-plane M(H) curves of sample S5 and S8, showing that the in-plane is the easy direction for both (demagnetization factor was taken into account in the out-of-plane curves).

TABLE 2. Temperature of the Verwey transition (T$_V$) and the width of the transition in the temperature axis (δT$_V$) determined from the derivative of the M$_{ZFC}$(T) fol all samples. Saturation magnetization (M$_S$, kA/m) for all samples obtained by extrapolating the M(H$^{-2}$) curve for H$^2$ → 0.

| Sample | S1 | S2 | S3 | S4 | S5 | S6 | S7 | S8 | S9 |
|---|---|---|---|---|---|---|---|---|---|
| T$_V$ (K) | 117.7 | 116.6 | 117.9 | 115.0 | 117.9 | 115.0 | 110.9 | 112.9 | 112.9 |
| δT$_V$ (K) | 9.9 | 9.5 | 9.1 | 11.7 | 12.5 | 12.1 | 11.5 | 12.0 | 9.6 |
| M$_S$ (kA/m) | 519 | 558 | 530 | 467 | 460 | 450 | 518 | 529 | 530 |

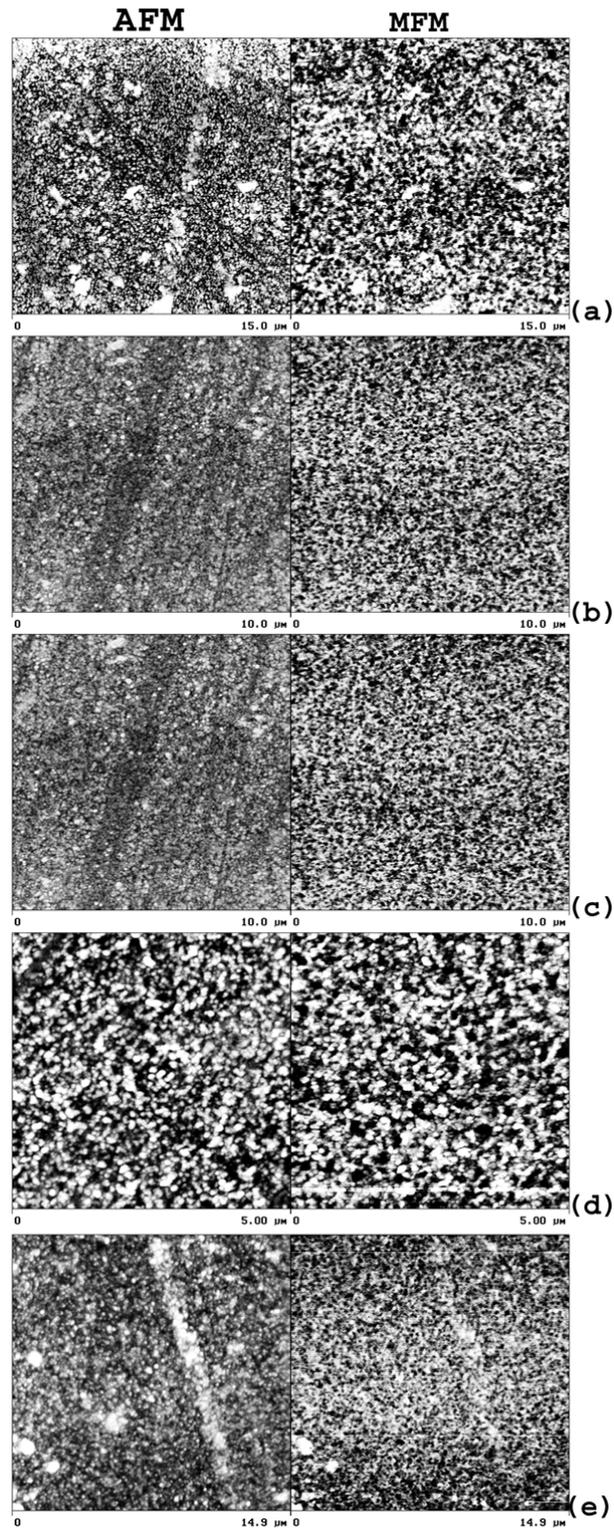

Figure 6. AFM and MFM images of samples (a) S2, (b) S5 and (c) S8 deposited on MgO (001) using both precursors, and AFM and MFM images of samples (d) S7 and (e) S9, deposited from precursor [Fe$^{II}$Fe$^{III}_2$(OBu$^t$)$_8$] on MgAl$_2$O$_4$ and Al$_2$O$_3$, respectively.

Figure 6-a-c displays the AFM and MFM images side-by-side for the samples S2, S5 and S8, respectively. These images indicate that there are differences in the patterns observed in MFM images and the morphology in the respective AFM ones. For samples S2 and S8, we observe that MFM pattern seems to be smaller than the grains observed in the AFM image, while for sample S5 we observed the opposite

situation: larger pattern in MFM image than in the AFM one. Thus, in the last sample, the magnetic domain probably incorporate more than one grain and in samples S2 and S8 one grain should present more than one magnetic domain. The AFM/ MFM images of samples S7 and S9 deposited on $MgAl_2O_4$ and $Al_2O_3$, respectively (Figures 6-d-e), showed no major differences with sample S8 (deposited on MgO). This result demonstrates that the final magnetic structure has a stronger dependence on the precursor than on the substrate, in agreement with the results from M(T) and M(H) data.

### *3.2 – Transport and Magneto-Transport*

The resistivity curves as function of temperature ($\rho(T)$, with 90 K < T < 300 K, excepting for samples S2, with 120 K < T < 300 K) of all samples show a continuous increase with decreasing the temperature for all samples, as expected for the magnetite. Figure 7-a displays the plot of $\ln(\rho(T))$ *vs.* 1/T for samples S1, S2 and S3 (deposited from precursor $[Fe^{II}Fe^{III}_2(OBu^t)_8]$ at 363 K) and sample S8 (precursor $[Fe^{II}Fe^{III}_2(OBu^t)_8]$ at 357 K on MgO), showing a linear behaviour and indicating a thermally-activated transport mechanism:

$$\rho(T) = \rho(0)\exp(E_a/k_B T),  \quad \text{eq. (2)}$$

where $\rho_0$ is the resistivity for $k_B T \gg E_a$ and $E_a$ is the activation energy. The resistivity values of this set of samples are lower with increasing the thickness of the films. For samples S1, S3 and S8, a discontinuity is observed for $T < T_V$ as consequence of the charge ordering below the Verwey transition.

For samples S4, S5 and S6 (deposited from precursor $[Fe^{III}_2(OBu^t)_6]$ at 354 K) ,and samples S7 and S9 (precursor precursor $[Fe^{II}Fe^{III}_2(OBu^t)_8]$ at 357 K on $MgAl_2O_4$ and on $Al_2O_3$, respectively), the curves $\ln(\rho(T))$ *vs.* 1/T do not present a linear behaviour, being closer to the linearity when plotted as function of $1/T^{1/2}$ (see figure 7-b).

These two distinct thermal behaviours observed for the $\ln(\rho(T))$ curves of our films indicate a difference in the fundamental mechanism of the charge transport in these systems. In a polycrystalline film, the thermal dependence of the resistivity with $1/T^{1/2}$ is expected, since the charge transport is dominated by the scattering of grain boundaries [6,18]. In this case, the resistivity depends majority from the grains size, which is associated preferentially to the synthesis conditions and not with the thickness of the film. In fact, we observe that samples S4, S5, S6, S7 and S9 present similar values of resistivity for all the temperature range measured, independently from the thickness of each sample.

For Samples S1, S2, S3 and S8, the linear dependence of $\ln(\rho(T))$ *vs.* 1/T supports the thermally-activated mechanism and suggests that spin-polarized transport through the anti-phase-boundaries (APBs) is the major mechanism. Accordingly, the anti-ferromagnetic coupling of APBs [29] will act as a scattering center for the fully spin-polarized electrons. In our highly crystalline, orientated and granular films, the presence of APBs is probably associated to the correlation between the morphological (grain size) and magnetic (domain size) characteristics lengths. For samples S1, S2, S3 and S8 the domain size seems to be smaller than the crystalline one. It is expected that the APBs density decreases with increasing the film thickness [17], and therefore the resistivity should decrease for thicker films. In our samples this characteristic can be clearly observed, as pointed out in figure 7-b.

From the linear fit of the data presented in figure 7-b with eq. (2), we determined the values of $E_a$ and $\rho_0$ for samples S1, S2, S3 and S8 (Table III). As expected $\rho_0$ decreases with increasing the thickness, varying from $0.3 \times 10^{-3}$ to $2.2 \times 10^{-3}$ $\Omega$.cm, in agreement with those ones obtained for epitaxial $Fe_3O_4$ films with 50 and 200 nm [17,29]. While the value of $E_a$ are almost constant among these samples, varying from 70 meV for sample S1 to 75 meV for sample S8, close to the values observed in the literature for epitaxial films of magnetite [2,4].

In general, the resistivity of films whose the scattering in the grain boundaries are dominant present for the polycrystalline films, where this mechanism is dominant, is larger than that of epitaxial ones [6,9,18,30]. Our results presented in figures 7-a and 7-b agrees with this prediction, with samples S4, S5, S6, S7 and S9 presenting a resistivity about 10 or 100 times greater than samples S1, S2, S3 and S8 for any temperature range. In Table III, we present the values of $\rho$ at T = 295 K ($\rho$(295 K)): for example, $\rho$(295 K) for sample S8 (thickness of 250 nm and transport dominated by APBs) is 10 times smaller than that of sample S9 (thickness of 360 nm and transport dominated by grains boundaries).

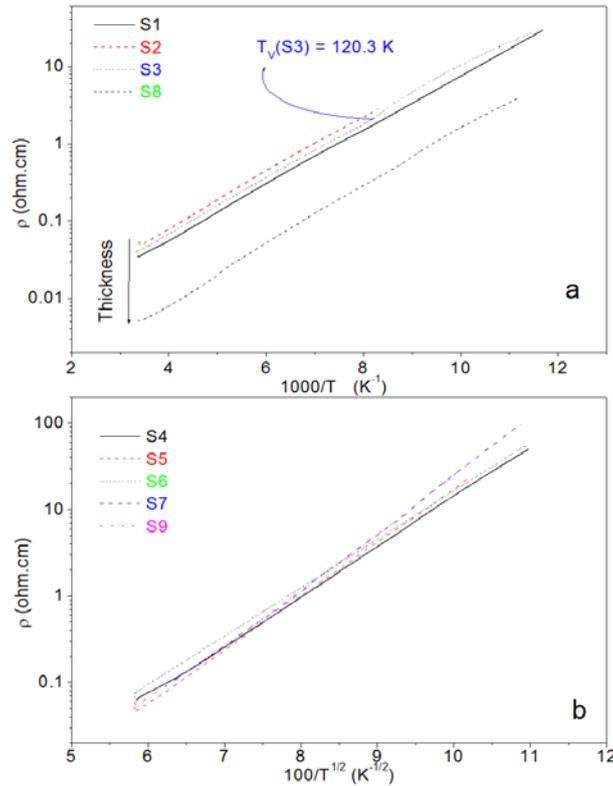

Figure 7. Figure 7-a shows the plot of $\ln(\rho(T))$ *vs.* 1/T for samples S1, S2 and S3 (deposited from precursor $[Fe^{II}Fe^{III}_2(OBu^t)_8]$ at 363 K) and sample S8 (precursor $[Fe^{II}Fe^{III}_2(OBu^t)_8]$ at 357 K on MgO). Figure 7-b shows the curves $\ln(\rho(T))$ *vs.* $1/T^2$ curves of samples S4, S5 and S6 (deposited from precursor $[Fe^{III}_2(OBu^t)_6]$ at 354 K), and samples S7 and S9 (precursor precursor $[Fe^{II}Fe^{III}_2(OBu^t)_8]$ at 357 K on $MgAl_2O_4$ and on $Al_2O_3$, respectively).

From the results presented above, it is clear that the structure of the precursor plays a fundamental role on the resulting magneto-transport properties. While the transport mechanism observed in the S1, S2 and S3 group (i.e., those from the precursor $[Fe^{II}Fe^{III}_2(OBu^t)_8]$ and $T_{pre}$= 363 K) was dominated by APB´s, those films from the S2, S3 and S4 group (precursor $[Fe^{III}_2(OBu^t)_6]$ and $T_{pre}$= 354 K) presented a transport behaviour dominated by grains boundaries. The precursor temperature $T_{pre}$ also present some influence, since group (S7,S8,S9, from precursor $[Fe^{II}Fe^{III}_2(OBu^t)_8]$ and $T_{pre}$= 357 K) present both mechanisms: for sample S8 (on MgO) the APB´s are dominant while in samples S7 ($MgAl_2O_4$) and S9 ($Al_2O_3$) the grains boundaries dominate the transport phenomenon. This analysis agree with the pointed out above concerning the MFM images of the films.

All samples presented a similar dependence of the resistivity with the applied field (MR(H) curve, obtained according eq. (1)), as shown in figure 8-a and 8-b for the in-plane and out-of-plane measurements, respectively. The MR value at 1592 kA/m in the in-plane configuration varies from 1.7 % for sample S1 up to 3.6 % for sample S8. For the out-of-plane configuration, we observe that the values are slightly smaller: 1.4 % and 3.1 % for sample S1 and S8, as shown in Table III. In the low field region (H < 398 kA/m), in-plane and out-of-plane MR(H) curves present linear and parabolic behaviour, respectively, while for H < 795 kA/m both curves shown a linear behaviour. This difference in the MR(H) curves at low-fields was observed in both polycrystalline [6,18,30] and epitaxial [17,29] films. For polycrystalline films, the linear and quadratic behaviour at low fields for in-plane and out-of-plane curves can be explained in terms of the tunnelling of the spin-polarized electrons through inter-grain anti-ferromagnetic boundaries [18]. In epitaxial films, these behaviours at low-fields can be explained in terms of a model based in the APBs and uniaxial anisotropy proposed by Eerenstein *et al* [29]. According to this model, the transport properties in the epitaxial $Fe_3O_4$ films are determined by the tunnelling of spin-polarized electrons through the thin and marked APBs (anti-ferromagnetic domain boundaries), adding the effects of a uniaxial anisotropy constant K for low applied fields. The conductivity σ for this kind of system calculated for a non-adiabatic limit is given proportional to $t^2$ α $cos^2\varphi_{AF}$, where $t^2$ is the transfer integral and $\varphi_{AF}$, is the angle between the moments in the antiferromagnetic boundary. The inclusion of the uniaxial anisotropy field ($H_{AN}$) defines two distinct regimes for $cos^2\varphi_{AF}$:

$$\cos^2\varphi_{AF} \propto \frac{(M_S H)^2}{K}, H < H_{an}$$

$$\cos^2\varphi_{AF} \propto M_S H - K, H > H_{an}. \qquad \text{eq. (3)}$$

Expanding the (field-independent) conductivity of bulk material in powers of a field-dependent term $\sigma_{AF}(H)$ (small perturbations), we obtain a field-dependent magnetoresistance given by:

$$MR(H) = \frac{\rho_H - \rho_0}{\rho_0} \approx -\left(\frac{d\sigma}{d\sigma_{AF}}\right)_0 \frac{d\sigma_{AF}}{\rho_0} \qquad \text{eq. (4)}$$

Therefore, in-plane and out-of-plane MR(H) curves must scale with the respective magnetization curves, in this case $(M/M_S)^2$ *vs.* H, for the low-field region. Figure 9 shows that the scaling between these curves is excellent for sample S8. In the high-field region (H > 795 kA/m) where the magnetization is almost saturated, the conductivity showed a linear dependence of the applied field.

As the grain boundaries act as scattering centers for the antiferromagnetically-copuled spin-polarized electrons, it is expected that samples S4, S5, S6, S7 and S9 display MR(H) curves that are similar to those measured for samples S1, S2, S3 and S8. Therefore the samples where grain boundaries or APBs dominate the transport properties will have different thermal dependence of the resistivity, but similar MH(R) curves.

Samples S7, S8 and S9 presented the largest MR values as consequence of their synthesis conditions. The higher MR values observed for the last set of samples could be associated with the larger density of APB´s and/or grain boundaries that was in turn originated from the faster growth rate of these samples. The different (faster) kinetics is likely to produce an increment of the 'seed' magnetite islands during the first stages of the deposition. Consequently, it could lead to an increment in the density of APB´s and grain boundaries, which is also reflected in the magnetic properties of the system such as the Verwey temperature (figure 4).

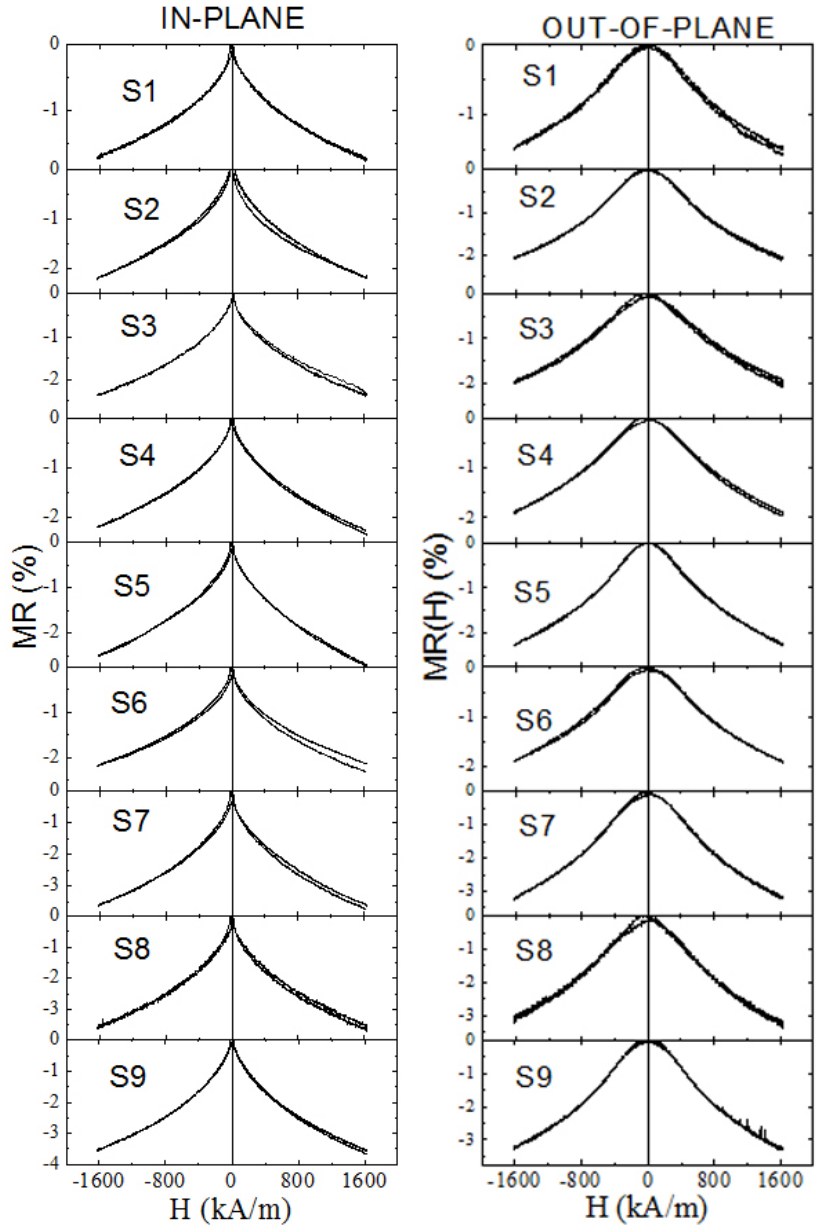

Figure 8. Figure 8-a and 8-b present the magneto-resistance curves with the applied field up to 1592 kA/m (MR(H), obtained according eq. (1)) for the in-plane and out-of-plane measurements, respectively, of all samples.

It is interesting to note the deviations observed in the contents of Fe and O with respect to the quantities expected for magnetite (relation Fe/O = 0.75), obtained from RBS data in these films. Indeed, for sample S2 and S7, the amount of Fe/O are 0.61 and 0.64, close to the value expected for maghemite (0.66). Although the presence of small amounts of maghemite could not be completely ruled out by the present data, it is worth to mention that the errors involved in RBS measurements and analysis (~ 5 %) could at least partially explain this deviation. However, the electrical behaviour of both gamma-$Fe_2O_3$ and $Fe_3O_4$ phases are completely different one from another. Maghemite is an insulator with resistivity values several orders of magnitude larger than those usually found for magnetite at room temperature. In our samples, the resistivity measured varied from 5 to 75 mΩ.cm, whereas maghemite values are usually within the 30 kΩ.cm range. Additionally, the films presented a clear Verwey transition, a signature of magnetite that is not present in maghemite (it is associated with the presence of $Fe^{2+}$ ion). It is interesting to mention that the magnetotransport effects observed for both mechanisms, AF-APBs and grain boundary scattering, are not expected for insulating systems such as maghemite. The above data make the possibility of maghemite to be present highly unlikely.

TABLE 3. Resistivity $\rho(T)$ and magnetoresistance MR(H) parameters for each sample. The activation energy $E_a$ and resistivity $\rho_0$ were obtained from the linear fitting of $\rho(T)$ curves using eq. (2); $\rho(295\ K)$ and MR(2T) were obtained directly from MR(H) curves.

| Sample | Transport Mechanism | $E_a$ (meV) | $\rho_0$ (mΩ.cm) | $\rho(295\ K)$ (mΩ.cm) | MR(2 T) % In-Plane | MR(2 T) % Out-Plane |
|---|---|---|---|---|---|---|
| S1 | APB's | 70 | 2.2 | 35.1 | -1.7 | -1.4 |
| S2 | APB's | 72 | 2.9 | 53.2 | -2.0 | -2.0 |
| S3 | APB's | 71 | 2.5 | 42.4 | -2.1 | -1.8 |
| S4 | Grain boundaries | - | - | 61.3 | -2.2 | -1.8 |
| S5 | Grain boundaries | - | - | 69.4 | -2.2 | -2.2 |
| S6 | Grain boundaries | - | - | 75.3 | -2.2 | -1.8 |
| S7 | Grain boundaries | - | - | 63.1 | -3.3 | -3.4 |
| S8 | APB's | 75.5 | 0.3 | 5.2 | -3.6 | -3.2 |
| S9 | Grain boundaries | - | - | 49.8 | -3.5 | -2.9 |

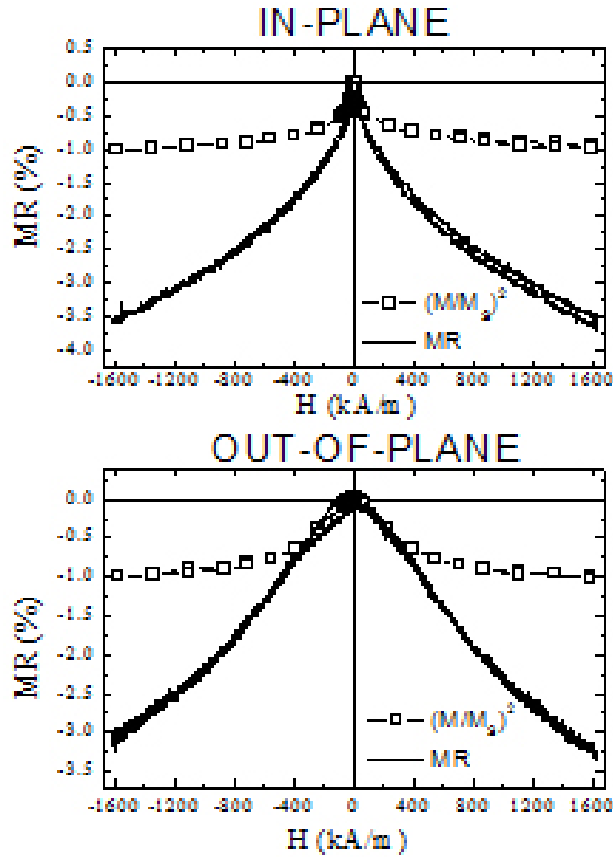

Figure 9. MR(H) and $(M/M_S)^2$ *vs* H curves of sample S8 in the in-plane and out-of-plane measurements.

## 4 – CONCLUSIONS

The first conclusion from the systematic analysis of the present series of magnetite films is the strong influence of the precursor type and temperature on both the resulting microstructural and magneto-transport properties. Regarding the $[Fe^{II}Fe^{III}_2(OBu^t)_8]$ precursor, the presence of both Fe(II) and Fe(III) centers in a single framework resulted in $Fe_3O_4$ films with higher $M_S$ and lower saturation field than those deposited from $[Fe^{III}_2(OBu^t)_6]$ precursor. By appropriate combination of the precursor kind, substrate and deposition temperatures $T_{pre}$, the observed scattering was dominated by either APB´s or grain boundaries. These results confirm the possibility to tune the magnetic and transport properties of magnetite films from CVD deposition by selecting appropriate deposition temperature and precursor type.

## LIST OF ABREVIATIONS

Activation Energy – $E_a$

Angle Between Magnetic Moments in the Antiferromagnetic Boundary – $\varphi_{AF}$

Antiferromagnetic Antiphase Boundaries – AF-APB

Antiphase Boundaries – APB

Applied Field – H

Atomic Force Microscope – AFM

Chemical Vapor Deposition – CVD

Coercive Field – $H_C$

Curie Temperature – $T_C$

Current – I

Deposition Time – DPT

Field-Cooling – FC

Full Width at Half Maximum – FWHM

Magnetic Force Microscopy – MFM

Magnetization – M

Magnetoresistance – MR

Molecular Beam Epitaxy – MBE

Precursor Temperature – $T_{pre}$

Pulsed Laser Deposition – PLD

Resistivity - $\rho$

Root Mean Square of Surface Roughness – rms

Rutherford Backscattering Spectroscopy – RBS

Saturation Magnetization – $M_S$

Substrate Temperature – $T_{sub}$

Temperature of Verwey Transition – $T_V$

Thickness of the Film – d

Uniaxial Anisotropy Field – $H_{AN}$

Voltage – V

Width of Verwey Transition – $\delta T_V$

X-ray Diffraction – XRD

Zero-Field-Cooling – ZFC